\documentclass[final,5p,times,twocolumn]{elsarticle}

\usepackage{mathptmx}
\usepackage{latexsym}
\usepackage{amssymb, amsmath, bm, physics}
\usepackage{graphicx,subfigure}
\usepackage{mathptmx}
\usepackage{subfigure}
\usepackage{flushend}
\usepackage[colorlinks,citecolor=blue,urlcolor=blue,linkcolor=blue]{hyperref}
\usepackage{aas_macros}
\usepackage{orcidlink}
\usepackage{float}
\usepackage{natbib}
\usepackage{times}
\begin{document}

\begin{frontmatter}




\title{Observational constraints on FLRW, Bianchi type I and V brane models}

\author{R. Jalalzadeh$^{1}$
\orcidlink{0000-0002-6110-3981}}
\ead{r.jalalzadeh@uok.ac.ir}
\author{S. Jalalzadeh$^{2,3}$ \orcidlink{0000-0003-4854-2960}}
\ead{shahram.jalalzadeh@ufpe.br}
\author{B. Malekolkalami$^{1}$ \orcidlink{0000-0002-5237-1730}}
\ead{b.malakolkalami@uok.ac.ir}

\author{Z. Davari$^{4}$ \orcidlink{}}
\ead{zahradavari@kias.re.kr}


\address{$^{1}$Department of Physics, University of Kurdistan, Pasdaran St., Sanandaj, Iran}
\address{$^{2}$Departamento de F\'{i}sica, Universidade Federal de Pernambuco,
Recife, PE, 50670-901, Brazil}
\address{$^{3}$Department of Physics and Technical Sciences, Western Caspian University, AZ 1001 Baku, Azerbaijan}
\address{$^{4}$Korea Institute for Advanced Study (KIAS), 85 Hoegiro, Dongdaemun-gu, Seoul, 02455, Korea}

\begin{abstract}
This study explores the compatibility of Covariant Extrinsic Gravity (CEG) with current cosmological observations. We employ the chi-square statistic and Markov Chain Monte Carlo (MCMC) methods to fit the Friedmann--Lemaître--Robertson--Walker (FLRW) and Bianchi type-I and V brane models to the latest datasets, including Hubble, Pantheon+ Supernova samples, Big Bang Nucleosynthesis (BBN), Baryon Acoustic Oscillations (BAO), and the structure growth rate, $f\sigma_8(z)$.  Parameters for FLRW universe consist $\left(\Omega^{\text{(b)}}_0, \Omega^{\text{(cd)}}_0, \Omega^{\text{(k)}}_0, H_0, \gamma, \sigma_8\right)$, while for the Bianchi model are $\left(\Omega^{\text{(b)}}_0, \Omega^{\text{(cd)}}_0, \Omega^{{(\beta)}}_0, H_0, \gamma, \Omega^{(\theta)}_0, \sigma_8\right)$. {By comparing our models to observational data, we determine the best values for cosmological parameters. For the FLRW model, these values depend on the sign of $\gamma$ (which gives the time variation of gravitational constant in Hubble time unit): $\gamma > 0$ yields $\gamma =0.00008^{+0.00015}_{-0.00011}$, and $\Omega^{\text{(k)}}_0=0.014^{+0.024}_{-0.022}$ and $\gamma < 0$ leads to $\gamma =-0.0226^{+0.0054}_{-0.0062}$, and $\Omega^{\text{(k)}}_0=0.023^{+0.039}_{-0.041}$. It should be noted that in both cases  $\Omega^{\text{(k)}}_0>0$, which represents a closed universe.
Similarly, for the Bianchi type-V brane model, the parameter values vary with the sign of $\gamma$, resulting in $\gamma = 0.00084^{+0.00019}_{-0.00021}$, $\Omega^{(\beta)}_0 =0.0258^{+0.0052}_{-0.0063} $, and $\Omega^{\theta}_0(\times 10^{-5} ) = 4.19^{+0.67}_{-0.75}$ (as with the density parameter of stiff matter) for $\gamma > 0$, and $\gamma = -0.00107^{+0.00019}_{-0.00020}$, $\Omega^{(\beta)}_0 = 0.0259^{+0.0050}_{-0.0062} $, and $\Omega^{\theta}_0(\times 10^{-5} )  = 4.17^{+0.91}_{-0.98}$ for $\gamma < 0$. In both 
cases $\Omega^{(\beta)}_0>0$, which represents the Bianchi type-V, because in the Bianchi type-I, $\beta=0$.} Subsequently, utilizing these obtained best values, we analyze the behavior of key cosmological parameters.
\end{abstract}

\begin{keyword}
Brane gravity\sep Bianchi cosmology\sep FLRW cosmology\sep Observational cosmology
\end{keyword}

\end{frontmatter}

\section{Introduction}\label{sec:int}
We are currently experiencing a golden age of cosmology. The unprecedented volume and accuracy of observational data allow for more precise analysis of cosmological phenomena. For instance, Type Ia supernovae (SNeIa) provide valuable information about the universe across the redshift scale $0.001 \leq z \leq 2.3$. Baryon Acoustic Oscillations (BAO) data offer insights within the range of $0.1 \leq z \leq 2$, while Cosmic Microwave Background (CMB) data extend our understanding to a redshift of $z=1100$. To gain a deeper understanding of the early universe, we should utilize observational data related to the formation of structures and the inflation era. When more observational data are used to constrain a cosmological model, it becomes more likely to be accepted and provides more accurate estimates of cosmological parameters.

The standard model of cosmology demonstrates the greatest compatibility with observational data and can explain a wide range of cosmological phenomena. It accounts for the universe's accelerating expansion, the CMB's power spectrum, the distribution of galaxies in large-scale structures, and the effects of gravitational lensing. Despite these successes, the model faces several fundamental challenges, including boundary conditions and initial singularity problems \cite{Jalalzadeh:2022bgz}, coincidence and fine-tuning issues~\citep{Velten:2014nra, Tian:2019enx, Jalalzadeh:2022dlj}, as well as observational challenges related to the nature of dark matter and dark energy~\citep{Pan-STARRS1:2017jku, Planck:2018vyg}. Additionally, there are cosmological tensions such as the Hubble tension and the amplitude of density fluctuations tension~\citep{Poulin:2022sgp}.

Theoretical models featuring large extra dimension(s) have gained significant attention in recent years. These models propose that the observable universe is embedded as a brane within a higher-dimensional space-time bulk. The concept of extra dimensions was first introduced by Kaluza and Klein in their seminal works published in 1921 and 1926, respectively \cite{Kaluza:1921tu, 1926ZPhy95K}. Their groundbreaking proposals suggested the existence of compact spatial dimensions beyond the familiar three dimensions of space and one dimension of time, revolutionizing theoretical physics. One notable model is the Space-Time-Matter (STM) theory presented by Wesson \cite{wesson-1998, Doroud:2009zza, Jalalzadeh:2008xu}, where the fifth dimension is non-compact, and the geometry of this extra dimension induces matter fields. Various classifications of brane universes have been proposed, including those by Arkani-Hamed, Dimopoulos, and Dvali, which aim to solve the hierarchy problem \cite{Arkani-Hamed:1998jmv, Arkani-Hamed:1998sfv}. Additionally, the Randall-Sundrum model developed a warped brane world to localize gravity \cite{Randall:1999vf}, and another braneworld model with extra dimensions of infinite volume provides a solution to the recent accelerated expansion of the universe \cite{Dvali:2000hr}.

Despite the many successes of these different brane models, they also face significant shortcomings. For example, all of these models rely on the junction condition to connect with bulk space, but different junction conditions can lead to diverse physical outcomes~\cite{Battye:2001pb}. Additionally, the compression of internal space can cause issues with the masses of fermion in nonzero modes and pose problems with vacuum stability~\citep{Arkani-Hamed:1999ylh}. Furthermore, challenges related to the thickness of these models present additional complications~\cite{Dzhunushaliev:2009va}.

In the CEG braneworld model \cite{Jalalzadeh:2013wza, Rostami:2015ixa, Jalalzadeh:2023upb} our $4D$ universe is embedded in a bulk space with an arbitrary number of non-compact extra dimensions. Consequently, the junction condition is not applicable. The features of this model can be summarized as follows (for details, see Ref. \cite{Jalalzadeh:2023upb}):
\begin{enumerate}
    \item Similar to the original Kaluza--Klein theories, the Yang--Mills Fields have a geometric origin. Like their predecessors, these theories explore the intricate relationship between geometry and physical phenomena, providing a holistic perspective on the underlying mechanisms at play in the universe. 
    \item The brane has a thickness proportional to the muon Compton wavelength, signifying a nuanced interplay between spatial dimensions and fundamental particles. 
    \item In the induced gravitational field equations, an extra geometric term, $Q_{\mu\nu}$ in Eq. (\ref{eq1}),  appears which could explain the present acceleration of the universe. This term is referred to as geometric dark energy (GDE). It is noteworthy that this geometric term, as demonstrated in Refs. \cite{Jalalzadeh:2005ax, Jalalzadeh:2004uv}, plays an essential role (acting as an induced potential) in the Klein-Gordon equation for particles confined to the brane.
    \item The energy density of GDE at the present epoch is determined by the number of extra dimensions $d$, $4D$ gravitational constant $G_N$, the fine structure constant $\alpha$, and the muon mass, $m_\mu$, given by $\rho^\text{(GDE)}_0=g(d)G_Nm_\mu^6/\alpha^2$. This relationship not only sheds light on the origins of Zeldovich's \cite{Zeldovich:1967gd} and Weinberg's \cite{Weinberg:1972kfs} renowned empirical formulas regarding vacuum energy and fundamental particle mass but also paves the way for a deeper comprehension of the cosmic landscape and its underlying principles. By intertwining geometry, field theory, and cosmology, these theories offer a unified framework that enriches our understanding of the universe's intricate tapestry and the forces that govern its evolution over vast cosmic scales.
    \item Due to its dependence on the extrinsic curvature radii, the $4D$ gravitational `constant' is generally variable. Within the isotropic and homogeneous cosmic framework, it may vary with time.
    \item Though these extra dimensions are non-compact, the CEG model predicts 22 extra dimensions, precisely as the original bosonic string theory did.
\end{enumerate}

{In the present article, we opt for the number of extra dimensions $d=22$, as determined in~\citep{Jalalzadeh:2023upb}, in which it was demonstrated that the CEG aligns with cosmological observational data when $d=22$. The non-compact nature of these additional dimensions is a key feature highlighted by the CEG model, which specifically forecasts a $26D$ bulk space, mirroring the foundational concepts of the original bosonic string theory. It is interesting to note that while the contemporary superstring theory is situated within 10 or 11 dimensions, there exists a hypothesis suggesting that all string theories could potentially be encompassed within the framework of the bosonic string theory operating in 26 dimensions. This particular proposition was initially put forth by researchers like Freund \cite{Freund:1984nd}, Casher et al. \cite{Casher:1985ra}, Englert et al. \cite{Englert:1986na}, and Nojiri \cite{Nojiri:1987ix, Nojiri:1986hp}. }

In most cosmological models, unknown parameters are inferred by aligning the model with observational data. However, given the uncertainties and errors inherent in such data, achieving a perfect match between the model and observations is unattainable even if the model is theoretically accurate. Therefore, it is crucial to calculate parameter errors and delineate confidence level contours relative to observational data.

Here, we use geometric observational data related to the cosmic background including 37 points of Hubble Observations data~\citep{Farooq:2016zwm}, a comprehensive set of 1701 point data from the latest supernova data called pantheon$+$~\cite{Scolnic:2021amr}, 1 point of BBN observational data~\cite{Serra_2009}, 9 points of BAO observational data~\cite{Camarena:2018nbr}, as well as 18 data points of f$\sigma_8(z)$ structural formation data~\citep{Nesseris:2017vor}. These data are utilized to estimate both cosmological parameters ($\Omega^{\text{(m)}}_0$, $\Omega^{\text{(k)}}_0$, $H_0$, $\sigma_8$) and model parameters ($\Omega^{(\beta)}_0$, $\gamma$, $\Omega^{(\theta)}_0$).

The structure of this study is as follows: in the section \ref{sec: dynamic}, we provide a summary of the formulas and results obtained in \cite{Jalalzadeh:2013wza, Jalalzadeh:2023upb}. In the section \ref{sec: static}, we employ statistical analysis to measure the free parameters of models using the MCMC method~\cite{Lewis:2002ah}. Finally, we analyze and review the results obtained in the sections \ref{sec: analyze} and \ref{concl}.

\section{ The dynamics of the FLRW, Bianchi type-I and V branes} \label{sec: dynamic}
To examine the FLRW and Bianchi type-I and V metrics in the context of CEG, we review the essential equations of the models. As outlined in \cite{Jalalzadeh:2013wza, Jalalzadeh:2023upb}, the induced gravitational field equations for a brane universe with thickness $l$ embedded in the bulk space with $d$ extra dimensions and the radius of curvature $L$ are given by: 
\begin{equation}
\label{eq1}
{G}_{\alpha\beta}= -Q_{\alpha\beta}+8\pi G_NT_{\alpha\beta},~~~~\alpha,\beta=0,...,3,
\end{equation}
where $G_{N}$ is the induced $4D$ Einstein tensor, and
\begin{multline}\label{eq5}
Q_{\alpha\beta}={{K}}_{\alpha}^{\,\,\,\,\eta a}{{K}}_{\beta\eta a}-{{K}}^a{{K}}_{\alpha\beta a}- \frac{1}{2}{{g}}_{\alpha\beta}({{K}}^{\mu\nu
a}{{K}}_{\mu\nu a}\\-{{K}}_a{{K}}^a),~~~~~a=4,...,D,
\end{multline}
is a conserved quantity ($\nabla^\beta Q_{\alpha\beta}=0$) induced by the extrinsic curvature, $K_{\mu\nu a}$ of the brane. 

Note that in $5D$ brane gravity ($D=5$), Israel's junction condition relates the extrinsic curvature to the energy-momentum tensor of the confined matter fields and $Q_{\mu\nu a}$ in Eq. (\ref{eq5}) gives us the quadratic contribution of the matter fields in the Einstein Eq. (\ref{eq1}).\\ 
Also in SMS braneworld model \cite{Shiromizu:1999wj}, the Codazzi equation 
\begin{equation}
    \label{sh1}
    \nabla_\beta K_a-\nabla_\alpha K^\alpha_{\,\,\beta a}=0,
\end{equation}
gives the conservation of the confined matter fields $T_{\mu\nu}$ \cite{Shiromizu:1999wj}. Conversely, in the CGE braneworld model, the extrinsic curvature is an independent variable due to the absence of Israel's junction condition, and its components may be (partially) determined by the Codazzi Eq. (\ref{sh1}). 

Also, in the CGE model, the $4D$ gravitational `constant' depends on the extrinsic curvature radius, $L$, and $D$-dimensional fundamental scale $M_D$ by
\begin{equation}
    \label{sh2}
    \frac{1}{16\pi G_N}=\frac{\pi^{\frac{d}{2}-1}}{16\Gamma(\frac{d}{2}+1)}L^dM_D^{d+2},
\end{equation}
where curvature radius is the minimum of the extrinsic curvature eigenvalues, $L=\min\{L_a \}$, and is obtained by
\begin{equation}
    \label{sh3}
    \det(g_{\mu\nu}-L^aK_{\mu\nu a})=0.
\end{equation}
In the cosmological setting, one can rewrite Eq. (\ref{sh2}) in the following form
\begin{equation}\label{6}
    G_N=G_0\left(\frac{L}{L_0}\right)^{-d},~~~~\frac{\dot G_N}{G_N}=-d\frac{\dot L}{L},
\end{equation}
where $G_0$ is the $4D$ gravitational constant at the present epoch, and the second relation is obtained from the time derivative of the first equation.

As it is mentioned in \cite{Jalalzadeh:2013wza}, there are various restrictions on how much $G_N$ (the gravitational constant) can locally change, including constraints from the binary pulsar $PSR_{1913}+ 16$. These limitations on $G_N$ directly correlate to the Hubble parameter \cite{damour1988limits, hellings1989experimental}. This means that the following equation holds
\begin{equation}\label{eq10}
    \frac{\dot G_N}{G_N} =\gamma H,~~~~|\gamma|<1.
\end{equation}
Substituting (\ref{6}) into (\ref{eq10}) gives us
\begin{equation}
    \label{sh5}
    \frac{\dot L}{L}=-\frac{\gamma}{d}H.
\end{equation}

Generally, the eigenvalues of the extrinsic curvature along the extra dimension in Eq. (\ref{sh3}) depend on space-time coordinates. Consequently, the $4D$ gravitational `constant' in Eq. (\ref{sh2}) is not a real constant. As a result, the Bianchi identity and the conservation of $Q_{\mu\nu}$ give us the following conservation of the confined matter fields
\begin{equation}\label{eq6}
  \nabla_\beta  \big(G_N T_{\alpha \beta}\big)=0.
\end{equation}

\subsection{FLRW brane cosmology}
First, we review the homogeneous and isotropic FLRW brane world as described in \cite{Jalalzadeh:2013wza}. The metric of an FLRW space-time is defined by
\begin{equation}\label{eq7}
    ds^2=-dt^2 + a(t)^2 \left(\frac{dr^2}{1-kr^2}+r^2d\Omega ^2 \right),
\end{equation}
where $k$ is the spatial curvature, and $k =-1, 0, 1$ 
represents an open, flat, and closed world, respectively. To align with the metric symmetries, the energy-momentum tensor and $Q_{\mu\nu}$ have the following form \cite{Jalalzadeh:2013wza} 
\begin{equation}
  \begin{split}\label{eq8}
      T_{\mu \nu}&=(\rho +p)u_\mu u_\nu +pg_{\mu \nu},~~~~u_\mu=-\delta_\mu^0,\\
       Q_{\mu\nu}&=-8\pi G_N\left\{(\rho^{\text{(GDE)}} +p^{\text{(GDE)}})u_\mu u_\nu +p^{\text{(GDE)}}g_{\mu \nu}\right\},
\end{split}      
  \end{equation}
  where $u^\mu$ is the component of the cosmic fluid velocity, $\rho$ and $p$ are the energy density and the pressure of the cosmic perfect fluid, and $\rho^{\text{(GDE)}}$ and $p^{\text{(GDE)}}$ are the energy density and the pressure associated to geometric dark energy. 
  
  As explained in detail in Ref. \cite{Jalalzadeh:2013wza}, by obtaining the components of the extrinsic curvature from Codazzi Eq. (\ref{sh1}) and utilizing Eq. (\ref{sh5}), one can find the components of  $Q_{\alpha \beta}$ defined by Eq. (\ref{eq5}). 
By inserting these components into Einstein's field Eqs. (\ref{eq1}), the Friedmann equations for the FLRW brane are obtained as follows:
\begin{subequations}\label{eq12}
    \begin{align}
& H^2+\frac{k}{a^2}=\frac{8\pi G_N}{3}\left(\rho +\rho^{\text{(GDE)}} \right),\\
 &\frac{\ddot a}{a}=-\frac{4\pi G_N}{3}\left(\rho + \rho^{\text{(GDE)}} +3p+3p^{\text{(GDE)}}\right),
    \end{align}
\end{subequations}
where
\begin{equation}\label{eq13}
    \begin{split}
    \rho^\text{(GDE)}&=\frac{3d}{8\pi G_NL^2}=\frac{3d}{8\pi G_0L_0^2}\left(\frac{a}{a_0}\right)^{\left(\frac{2}{d}-1\right)\gamma},\\
    p^\text{(GDE)}&=-\frac{3d}{8\pi G_0L_0^2}\left(1+\frac{2\gamma}{3{d}}\right)\left(\frac{a}{a_0}\right)^{\left(\frac{2}{d}-1\right)\gamma}.
    \end{split}
\end{equation}

 The equation of state parameter for the GDE, $\omega_{\text{GDE}}$, is defined as 
\begin{equation}\label{eq14}
\omega_{\text{GDE}}=\frac{p^{\text{(GDE)}}}{\rho^{\text{(GDE)}}}=-\left(1+\frac{2\gamma}{3d}\right).
\end{equation}
Also, Eq. (\ref{eq6}) gives us the modified conservation of energy-momentum of the perfect fluid
\begin{equation}\label{eq15}
   \frac{\dot \rho}{\rho}=-3 (1+\omega)H-\frac{\dot G_N}{G_N}=\left\{ -3(1+\omega)+\gamma\right\}H, 
 \end{equation}
where we used Eq. (\ref{eq10}) in the second equality.

Let us assume the perfect cosmic fluid comprises two components: cosmic radiation and cold matter including baryonic and dark matters. Based on this assumption, we can define density parameters for different components at the present epoch. These parameters include: $\Omega^{\text{(GDE)}}_0$ for GDE, $\Omega^{\text{(m)}}_0$ for cold dust, $\Omega^{\text{(r)}}_0$ for radiation, and $\Omega^{\text{(k)}}_0$ for the curvature
\begin{eqnarray}\label{eq17}
    \begin{array}{cc}
      \Omega^{\text{(GDE)}}_0=\frac{8\pi G_0\rho^{\text{(GDE)}}_0}{3H_0^2},&
      \Omega^{\text{(m)}}_0=\frac{8\pi G_0\rho_0^{\text{(m)}}}{3H_0^2},\\
      \Omega^{\text{(r)}}_0=\frac{8\pi G_0\rho_0^{\text{(r)}}}{3H_0^2},& \Omega^{\text{(k)}}_0=\frac{8\pi G_0\rho_0^{\text{(k)}}}{3H_0^2}.
    \end{array}
\end{eqnarray}

Utilizing the above definitions and defining dimensionless Hubble parameter as $E(z)= \frac{H(z)}{H_0}$, we can rewrite the Friedmann Eqs. \eqref{eq12} to obtain the Hubble parameter and the deceleration parameter, $q$, as follows
\begin{multline}\label{eq18}
E^2=\Omega^{\text{(r)}}_0(1+z)^4+\Omega^{\text{(m)}}_0(1+z)^3+\\ \Omega^{\text{(GDE)}}_0(1+z)^{3(\omega_{\text{(GDE)}}+1)}+\Omega^{\text{(k)}}_0(1+z)^2,
  \end{multline}
  \begin{multline}\label{eq19}  
2qE^2=2\Omega^{\text{(r)}}_0(1+z)^4+\Omega^{\text{(m)}}_0(1+z)^3+\\(1+3\omega_{\text{(GDE)}})\Omega^{\text{(GDE)}}_0(1+z)^{3(\omega_{\text{(GDE)}}+1)}. 
\end{multline}
At the present epoch, the Eq. \eqref{eq18} is converted to 
 \begin{equation}\label{eq20}
\Omega_0^{\text{(m)}}+\Omega_0^{\text{(GDE)}}+\Omega_0^{\text{(r)}}+\Omega_0^{\text{(k)}}=1.
 \end{equation}
Also, using Eqs. \eqref{eq19}, \eqref{eq14}, and defining
\begin{equation}
    \begin{split}
      \Omega^{\text{(GDE)}}&=E^{-2}\Omega^{\text{(GDE)}}_0(1+z)^{3(\omega_{\text{(GDE)}}+1)},\\
      \Omega^{\text{(m)}}&=E^{-2}\Omega^{\text{(m)}}_0(1+z)^3,\\
      \Omega^{\text{(r)}}&=E^{-2}\Omega^{\text{(r)}}_0(1+z)^4,
    \end{split}
\end{equation}
we obtain the following equation for the decelerating parameter
\begin{equation}\label{eq21}
    q=\frac{1}{2}\Omega^{\text{(m)}}+\Omega^{\text{(r)}}-(1+\frac{\gamma}{d})\Omega^{\text{(GDE)}}.
\end{equation}
\subsection{Bianchi type-I and V cosmology}
The second braneworld, which has been investigated in Ref. \cite{Jalalzadeh:2023upb}, is the Bianchi type-I and V with the following metric
\begin{multline}\label{eq22}
      ds^2=-dt^2+a_1(t)^2dx^2+\\a_2(t)^2e^{2\beta x}dy^2+a_3(t)^2e^{2\beta x}dz^2,
  \end{multline}
which for Bianchi type-I, $\beta=0$ and for type-V, $\beta=1$. We define the average scale factor, $a(t)$
\begin{equation}\label{eq23}
a(t)=\Big(a_1(t)a_2(t)a_3(t)\Big)^\frac{1}{3},
\end{equation}
and the generalized mean Hubble’s parameter as
\begin{equation}\label{eq24}
    H=\frac{1}{3}(H_1+H_2+H_3).
\end{equation}
Similar to the FLRW model discussed in the preceding subsection, one can derive the components of $Q_{\alpha \beta}$ as defined by Eq. (\ref{eq5}) by determining the extrinsic curvature through the Codazzi equation (\ref{sh1}) and employing Eq. (\ref{sh5}). Subsequently, upon insertion of these components into Einstein's field Eqs. (\ref{eq1}), as it is shown in \cite{Jalalzadeh:2023upb}, the Friedmann equations can be obtained
\begin{multline}\label{eq25}
            H^2=\frac{8\pi G_0}{3}\Big\{\rho_0a^{-3(1+\omega)}+\\
        \rho^\text{(GDE)}_0a^{-3(1+\omega_\text{(GDE)})}\Big\}+
        \frac{\beta^2}{a^2}+\frac{B^2}{3a^6},
\end{multline}   
and
\begin{multline}\label{eq26}                \frac{\ddot a}{a}=-\frac{4\pi G_0}{3}\Big\{(1+3\omega)\rho_0a^{-3(1+\omega)}+\\(1+3\omega_\text{(GDE)})\rho^\text{(GDE)}_0a^{-3(1+\omega_\text{(GDE)})}\Big\}-\frac{2B^2}{3a^6},
\end{multline}
where $B$ is the constant of integration, for more details see~\cite{Jalalzadeh:2023upb}.

By defining the density parameters as follows
\begin{eqnarray}
    \label{eq27}
    \begin{array}{cc}
      \Omega^\text{(GDE)}_0=\frac{8\pi G_0\rho_0^\text{(GDE)}}{3H_0^2},&
      \Omega^\text{(m)}_0=\frac{8\pi G_0\rho_0^{\text{(m)}}}{3H_0^2},\\
      \Omega^{\text{(r)}}_0=\frac{8\pi G_0\rho_0^{\text{(r)}}}{3H_0^2},&
      \Omega^{(\theta)}_0=\frac{8\pi G_0\rho_0^{\theta}}{3H_0^2},\\
      \Omega^{{(\beta)}}_0=\frac{8\pi G_0\rho_0^{\beta}}{3H_0^2},&
    \end{array}
\end{eqnarray}
which $\rho_0^{(\theta)}=\frac{B^2}{8\pi G_0}$ and $\rho_0^{(\beta)}=\frac{3\beta^2}{8\pi G_0}$, we can rewrite Friedmann Eqs. (\ref{eq25}) and (\ref{eq26}) for the Bianchi type-I and V as follows
\begin{multline}\label{eq28}
E^2=\Omega^\text{(r)}_0(1+z)^4+\Omega^\text{(m)}_0(1+z)^3+\Omega^\text{(GDE)}_0(1+z)^{3(\omega_{\text{(GDE)}}+1)}+\\\Omega^{(\theta)}_0(1+z)^6+\Omega^{{(\beta)}}_0(1+z)^2,
  \end{multline}
  \begin{multline}\label{eq29}  
2qE^2=2\Omega^\text{(r)}_0(1+z)^4+\Omega^\text{(m)}_0(1+z)^3+\\(1+3\omega_\text{(GDE)})\Omega^\text{(GDE)}_0(1+z)^{3(\omega_{\text{(GDE)}}+1)}+4\Omega^{(\theta)}_0(1+z)^6.
\end{multline}

Finally, using Eqs.~\eqref{eq14} and 
\eqref{eq29} and defining
\begin{equation}
    \begin{split}
      \Omega^{\text{(GDE)}}&=E^{-2}\Omega^{\text{(GDE)}}_0(1+z)^{3(\omega_{\text{(GDE)}}+1)},\\
      \Omega^{\text{(m)}}&=E^{-2}\Omega^{\text{(m)}}_0(1+z)^3,\\
      \Omega^{\text{(r)}}&=E^{-2}\Omega^{\text{(r)}}_0(1+z)^4,\\
      \Omega^{(\theta)}&=E^{-2}\Omega^{(\theta)}_0(1+z)^6,\\
      \Omega^{(\beta)}&=E^{-2}\Omega^{(\beta)}_0(1+z)^2,
    \end{split}
\end{equation}
we obtain $q$ (the deceleration parameter) as below
\begin{equation}
    q=\frac{1}{2}\Omega^{\text{(m)}}+\Omega^{\text{(r)}}-(1+\frac{\gamma}{d})\Omega^{\text{(GDE)}}+2\Omega^{(\theta)}.
\end{equation}
 
\section{Statistical analysis of the models}\label{sec: static}
Our goal in the statistical analysis of the CEG model is to estimate the cosmic parameters and parameters related to the model by constraining it to both background (geometric data) and perturbation level (dynamic data of the growth rate) observations. Since these observational datasets are independent, the total probability is equal to the product of their probabilities. Thus, the probability function (the total likelihood function) is given by:
\begin{equation}
\mathcal{L}_{\text{tot}}(p)=\mathcal{L}_{\text{SN}} \cdot \mathcal{L}_{\text{BAO}} \cdot \mathcal{L}_{\text{OHD}} \cdot \mathcal{L}_{\text{BBN}} \cdot \mathcal{L}_{\text{fs}}.
\end{equation}

Since $\mathcal{L}_{\text{tot}} \simeq e^{-\frac{\chi^2_{\text{tot}}}{2}}$, we can write the total least squares function as follows
\begin{multline}\label{eq30}
     \chi^2_{\text{tot},2}(p)=\chi^2_{\text{BBN}}(p)+\chi^2_{\text{OHD}}(p)+\chi^2_{\text{SN}}(p)+\\ +\chi^2_{\text{BAO}}(p)+\chi^2_{\text{fs}} (p).
\end{multline}
Then, using the MCMC algorithm~\cite{Lewis:2002ah}, we constrain the vector of parameters ($p$) for the FLRW model, including $\left(\Omega^{\text{(b)}}_0, \Omega^{\text{(cd)}}_0, \Omega^{\text{(k)}}_0, H_0, \gamma, \sigma_8\right)$ and for Bianchi type-I and V including $\left(\Omega^{\text{(b)}}_0, \Omega^{\text{(cd)}}_0, \Omega^{(\beta)}_0, H_0, \gamma, \Omega^{(\theta)}_0, \sigma_8\right)$. 

To clarify, we write the total matter density parameter as $\Omega^{\text{(m)}}_0=\Omega^{\text{(cd)}}_0+\Omega^{\text{(b)}}_0$, which $\Omega^{\text{(cd)}}$ and $\Omega^{\text{(b)}}_0$ are cold dark matter density parameter and baryonic matter density parameter, respectively, and $\Omega^{\text{(r)}}_0=\Omega^{\text{(ph)}}_0+\Omega^{(\nu)}_0$ which $\Omega^{\text{(ph)}}$ and $\Omega^{(\nu)}_0$ are photon density parameter and neutrino density parameter, respectively. We derive the parameter values that best fit the observational data by minimizing the square function $(\chi^2_{\text{tot}})$ or maximizing the likelihood. To achieve this, we incorporate data from various sources including Big Bang Nucleosynthesis (BBN, 1 data point) \cite{Serra_2009}, Hubble evolution data (OHD, 37 data points) \cite{Farooq:2016zwm}, type Ia supernova data (SNIa, 1701 data points from Pantheon$+$ sample) \cite{Scolnic:2021amr}, baryon acoustic oscillation data (BAO, 9 data points) \cite{Camarena:2018nbr}, and 18 independent data points of the growth rate obtained from Redshift Space Distortions (RSD) in various galaxy surveys \cite{Nesseris:2017vor}. {It is noteworthy to highlight that, $\Omega^{\text{(k)}}_0$ and $\Omega^{(\beta)}_0$ are the free parameters of the FLRW and Bianchi models, respectively. It means that we obtain the value of these parameters using the constraining of models with the latest datasets. Therefore, in the FLRW model, we don't take a special value to $K$; in the Bianchi model, we don't take a special value to $\beta$. On the other hand, to categorize the available models, we use the different signs of $\gamma$ ($\gamma<0$, $\gamma=0$, and $\gamma>0$) in the FLRW and Bianchi models.}
The following is a brief overview of the datasets used in this study. 
\subsection{Hubble parameter observational data}

The Hubble parameter directly reflects the rate of expansion of the universe. The following two methods measure this parameter at a specific redshift. The first method, known as the cosmic chronometer, utilizes age changes in a series of galaxies to derive the Hubble parameter directly from $H(z)=\frac{-1}{1+z}\frac{dz}{dt}$~\cite{simon2005constraints, Moresco:2018xdr}. The second method involves measuring the Hubble parameter using the position of the BAO peaks, which allows for the measurement of the Hubble parameter in the radial direction using the equation $\int_{z_d}^\infty \frac{c_s(z) dz}{H(z)}$~\cite{anderson2014clustering, Marra:2017pst}.

Here, we utilize a standard set of 37 data points providing precise estimates of the Hubble parameter in the range $0.07 \leq z \leq 2.3$~\cite{Farooq:2016zwm}. Due to the independence of these observational data, we can define the chi-squared statistic as:
\begin{equation}\label{eq31}
 \chi^2_{\text{OHD}}(p)=\sum_{i=1} ^{37}\frac{\left[H_{\text{th}}(z_i,p)-H_{\text{ob}}(z_i)\right]^2}{\sigma^2_i},   
\end{equation}
where $p$ represents the vector of the model's free parameters, $H_{\text{th}}$ denotes the theoretical value of H(z) and $\sigma_i$  is the Gaussian error at the specific measurement $z_i$.


\subsection{Observational data of type I supernovae}

Standard candles, particularly type I supernovae, serve as crucial cosmological probes, with their intrinsic luminosity allowing for the measurement of relative distances. Various supernova datasets are available ~\cite{SupernovaCosmologyProject:2011ycw, Camarena:2018nbr}, among which the most recent is the Pantheon$+$ dataset~\cite{Scolnic:2021amr}. This dataset contains 1701 data points spanning the redshift range $0.001 \leq z_i \leq 2.26$, providing vital insights into the nature of the expanding universe. In the case of the Pantheon$+$ dataset,  we can define the chi-squared statistic as:
\begin{equation}\label{eq32}
     \chi^2_{\text{Pntheon}+}(p)=\Vec{D^T}.C_{\text{Pantheon}+}^{-1}.\Vec{D},
\end{equation}
where the quantity $\Vec{D}$ denotes the discrepancy between the observed apparent magnitudes $m_i$ of Type Ia supernovae (SNIa) and the anticipated magnitudes determined by the model, parameter $M$ signifies the absolute magnitude of SNIa.

In contrast, $\mu_{model}$ denotes the distance modulus the assumed cosmological model anticipated. The symbol $C_{\text{Pantheon+}}$ represents the covariance matrix accompanying the Pantheon$+$ dataset, encompassing systematic and statistical uncertainties. The distance modulus serves as a metric of the distance to an entity, specifically defined as
\begin{equation}
\begin{split}
    \mu_\text{model}(z_i)=&5\log_{10}\frac{d_{L}(z_i)}{Mpc}+25,\\
    d_L(z)=&c(1+z)\int^z _0\frac{dz}{H(z)},
    \end{split}
\end{equation}
where $d_L(z)$ is the luminosity distance. 

The Pantheon$+$ sample comprises 1701 data points, with 77 specifically associated with galaxies hosted by Cepheids within the low redshift interval of $0.00122 \leq z \leq 0.01682$. To break the degeneracy between $H_0$ and the $M$ of Type Ia, we define the modified vector $\Vec{D^{'}}_i$ as
\begin{equation}
    \Vec{D^{'}}_i =
\left\{
	\begin{array}{ll}
		m_i-M-\mu_i,  & \mbox{ } i\in \text{Cepheid hosts} \\
		m_i-M-\mu_\text{model}(z_i),  & \mbox{ } \text{otherwise}
	\end{array}
\right.
\end{equation}
where $m_i$ is the apparent magnitude and $\mu_i$ is the distance modulus of the ith SN. As a result of this modification, we can write the chi-square equation for SNe in 77 Cepheid host galaxies $(\chi^2_{\text{Cepheid}})$ and the remaining 1624 SNe $(\chi^2_{\text{SN}})$ as below
\begin{equation}
\begin{split}
\chi^2_{\text{Cepheid}}(p)=&\Vec{{D^{'}}}^T.C_{\text{Cepheid}}^{-1}.\Vec{D^{'}},\\
\chi^2_{\text{SN}}(p)=&\Vec{{D^{'}}}^T.C_{\text{SN}}^{-1}.\Vec{D^{'}}.
\end{split}
\end{equation}
The total likelihood will be
\begin{equation}
   \chi^2=\chi^2_{\text{Cepheid}}+\chi^2_{\text{SN}}
\end{equation}

\subsection{Observational data of Big Bang Nucleosynthesis}
Big Bang Nucleosynthesis involves the formation of heavy nuclei from hydrogen in the early times of the universe formation. The observed abundances of these primary nuclei serve as crucial constraints for various cosmological models and parameter measurements.  The predicted initial abundance of  $He^4$ is around $25\%$. The mass fraction of $He^4$ or $Y_{He}$ exhibits two compatible measurements derived from regression analyses in blue dense galaxies:  $Y_{He}=0.231 \pm 0.003$
And $Y_{He}=0.244 \pm 0.002$. Here, we adopt the value of $Y_{He}=0.244 \pm 0.002$. 
Additionally, recent measurements yield the weighted abundance of deuterium in the Lyman-$\alpha$ forest as  $\frac{D}{H}=(2.2 \pm 0.2) \times 10^{-5}$. We utilize this data to constrain the baryon density. Therefore, the chi-squared statistic for BBN can be expressed as~\cite{Serra_2009, burles2001big}
\begin{equation}\label{eq33}
    \chi^2_{\text{BBN}}(p)=\frac{[\Omega^{\text{(b)}}_0 h^2-0.022]^2}{0.002^2}.
\end{equation}
\subsection{Observational data of baryonic acoustic oscillations}

Baryon acoustic oscillations~\cite{beutler20116df, blake2011wigglez} refer to periodic density fluctuations arising from the interplay between gravity and pressure in the baryonic matter of the cosmic fluid.  BAO serves as a standard ruler for measuring cosmic distances. This phenomenon enables the constraint of the angular diameter distance and the Hubble parameter.

In this study, we employ the method outlined in \cite{Camarena:2018nbr}, which incorporates data points from various surveys including 6dFGS, SDSS-LRG, BOSS-MGS, BOSS-LOWZ, WiggleZ, BOSS-CMASS, and BOSS-DR12, to constrain the CEG model. We divide the dataset into two groups demonstrated in Table II and III in \cite{Camarena:2018nbr}.

The function $\chi^2_{\text{BAO},1}$ corresponds to the independent data listed in Table II \cite{Camarena:2018nbr}
\begin{equation}\label{eq34}
    \chi^2_{\text{BAO},1}=\sum_{i=1}^{2}\frac{[d_{z,i}-d_z(z_i)]^2}{\sigma_i^2},
\end{equation}
where the quantity of $d(z)$ is defined as follows
\begin{equation}\label{eq35}
    d_z(z)=\frac{r_s(z_d)}{D_v(z)},
\end{equation}
in which $D_v(z)$ is the isotropic angular diameter distance, and $r_s(z_d)$ is the comoving sound horizon distance in the photon decoupling epoch
\begin{equation}\label{eq36}
    r_s(z)=\int^\infty_z c_s(z)/H(z).
\end{equation}
Also, $c_s(z)$ represents the baryon sound speed, derived from the following equation
\begin{equation}\label{eq37}
     c_s(z)=\frac{1}{\sqrt{3+\frac{9\Omega^{\text{(b)}}_0}{4\Omega^{(ph)}_0(1+z)}}}.
\end{equation}

Similar to the approach introduced in \cite{Camarena:2018nbr}, we set the value of $\Omega^{(ph)}_0$ (photon energy density parameter) to $2.469 \times 10^{-5} h^{-2}$. $z_d$ represents the redshift of decoupling, determined using the fitting relation proposed in Ref. \cite{Hu:1995en} as
\begin{equation}\label{eq38}
    \begin{split}
        &z_d=\frac{1291(\Omega^{\text{(m)}}_0 h^2)^{0.251}\left(1+b_1(\Omega^{\text{(b)}}_0)^{b_2}\right)}{1+0.659(\Omega^{\text{(m)}}_0 h^2)^{0.828}},
        \\
        &b_1=\left(0.313(\Omega^{\text{(m)}}_0 h^2)^{-0.419}\right)(1+0.607(\Omega^{\text{(m)}}_0 h^2)^{0.674}),\\
        &b_2=0.238\left(\Omega^{\text{(m)}}_0 h^2\right)^{0.223}.
    \end{split}
\end{equation}

The second function $\chi^2_{\text{BAO},2}$  corresponds to the uncorrelated data outlined in Table III, as well as the correlated data of the Wiggle Z matrix detailed of \cite{Camarena:2018nbr}, defined as
\begin{equation}\label{eq39}
    \chi^2_{\text{BAO},2}=\{\alpha^*_i-\alpha^*(z_i)\}\Sigma_{\text{BAO},ij}^{-1}\{\alpha^*_j-\alpha^*(z_j)\},
\end{equation}
where the sum over the indices is implied, $\alpha^*(z)$ is given by
\begin{equation}\label{eq40}
   \alpha^*(z)=\frac{D_v(z)}{r_s(z_d)}r^{\text{fid}}_s,
\end{equation}
and $\Sigma^{-1}$ is the inverse matrix of the covariance matrix
\begin{equation}
    \Sigma_\text{Wiggle Z}=\begin{pmatrix}
6889 & -8961 & 21277 \\
 & 10201 & -13918 \\0
  &   & 7396   
\end{pmatrix}.
\end{equation}
The parameters, including $d_z$, $\sigma$, $\alpha^*$, and $r^{\text{fid}}_s$, for all mentioned surveys, are provided in Tables II and III of \cite{Camarena:2018nbr}.
\subsection{Observational data on Redshift Space Distortions}
Another important set of observational data, which strongly constrains cosmological models and helps differentiate between them, pertains to the growth rate of large-scale structures. We utilize 18 independent data points $f\sigma_8(z)$\cite{Nesseris:2017vor} to constrain the CEG model in FLRW and Bianchi type-I and V metrics. These observational data, extracted from various cosmic probes such as WiggleZ, THF, PSC, 2DF, VVDS, DNM, 6dFGRS, Boss, and 2MASS using the Redshift Spatial Distortion method (RSD), are compiled. Due to the independence of these observational data, the function $\chi^2_{\text{fs}}$ is defined as
\begin{equation}\label{eq47}
    \chi^2_{\text{fs}}(p)=\sum_{i=1}^{18} \frac{[(f\sigma_8)_{th}(z_i,p)-(f\sigma_8)_{ob}]^2}{\sigma^2_i}.
\end{equation}

In the following, we investigate the growth of matter density perturbation in the context of the CEG model. The evolution equation for matter perturbations in a fluid containing non-relativistic matter is given by \cite{Mehrabi:2015kta}
\begin{equation}\label{eq48}
 \ddot{\delta}+2H\dot \delta-4\pi G_N(t)\rho (t)\delta=0.
\end{equation}

By employing Eqs. \eqref{6}, \eqref{eq15}, and \eqref{eq48} and replacing the time derivatives with derivatives of the scale factor, we derive the following equation:
\begin{equation}\label{eq49}
\begin{split}
    \delta^{''}_m+\left(\frac{3}{a}+\frac{H^{'}}{H}\right)\delta^{'}_m=\frac{3}{2a^2}\Omega^{\text{(m)}}\delta_m,
    \end{split}
\end{equation}
To solve the above equation, we set \textbf{$a_i=0.001$} (initial scale factor, investigating the evolution of structure in a matter-dominated era), and for linear perturbations, we consider $\delta_{m, i} \propto a_i=1.6 \times 10^{-4}$. We employ the initial condition $\delta^{'}_{m, i}=\frac{\delta_{m, i}}{a_i}$, and numerically solve the equation up to the present time scale factor $a_0=1$~\cite{batista2013structure}.


\section{Analysis of the results}\label{sec: analyze}

\begin{figure*}[!htbp]
	\begin{center}
  \subfigure[]{\includegraphics[width=0.45\textwidth]{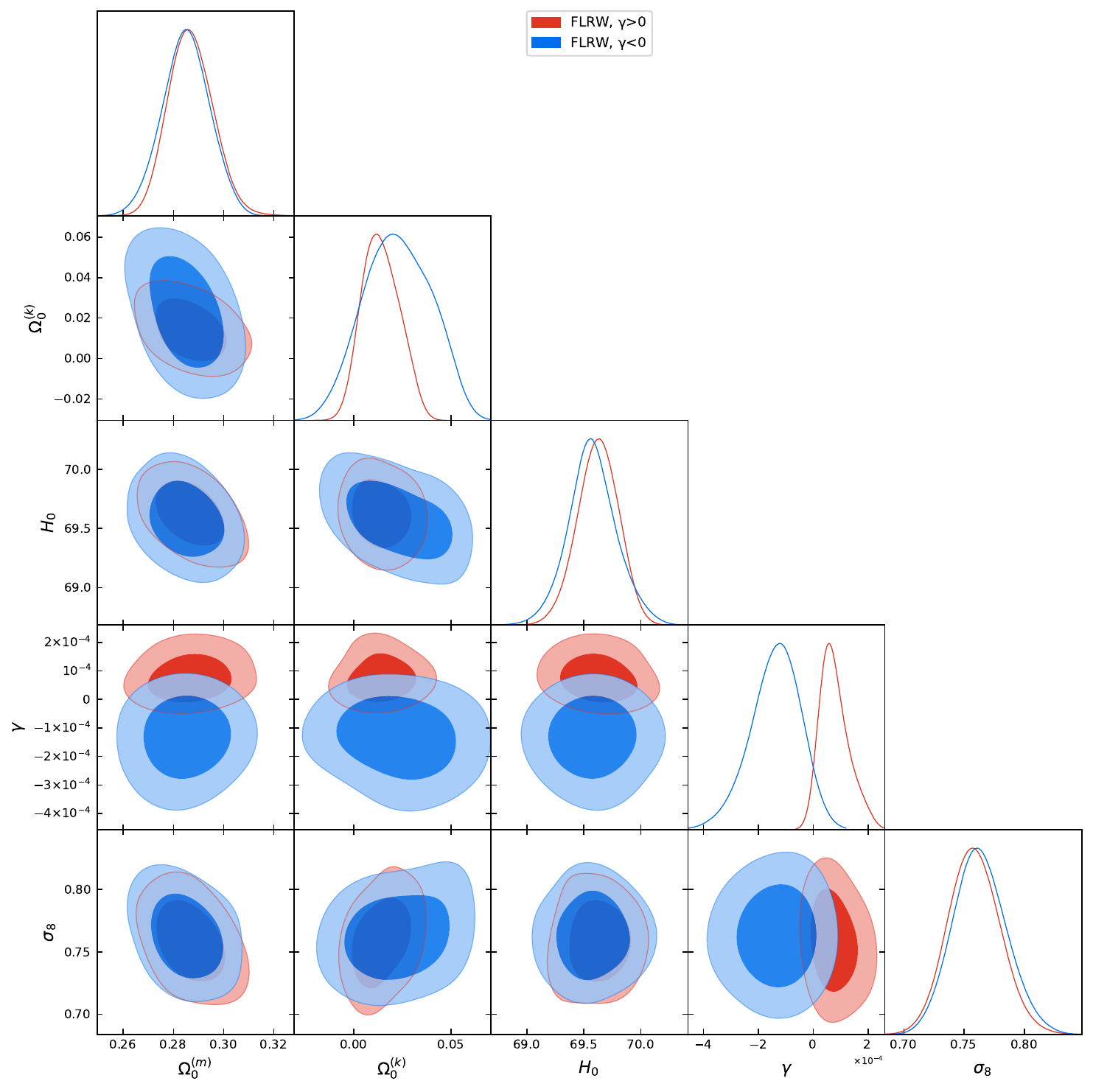}}\subfigure[]{\includegraphics[width=0.45\textwidth]{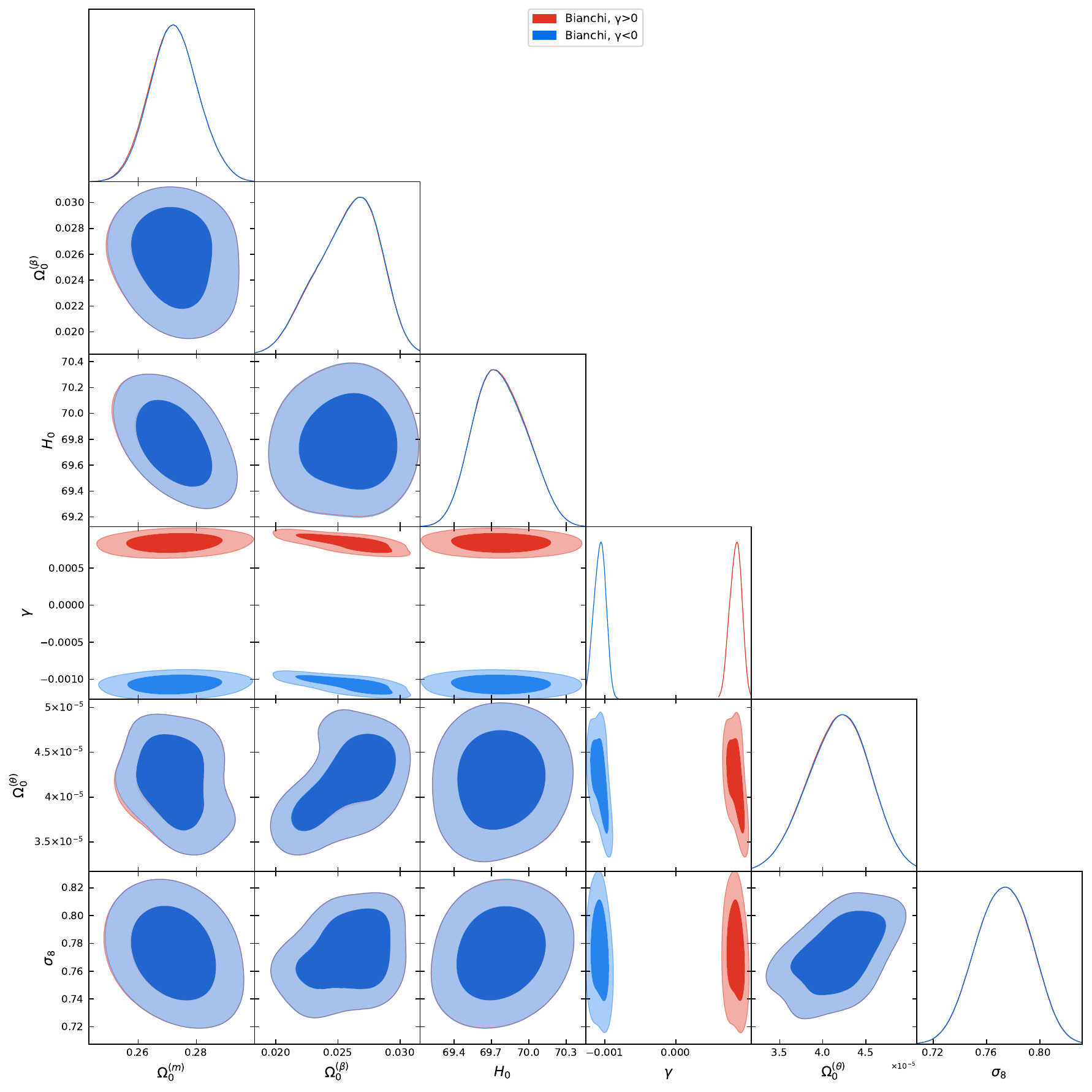}}
  
  \caption{\small a) The likelihood contours for FLRW model parameters in two cases $\gamma>0$ and $\gamma<0$ using the joint dataset: a level of confidence of $68.26\%$ ($1-\sigma$) and $95.44\%$  ($2-\sigma$), b) The likelihood contours for Bianchi type-I and V model parameters in two cases $\gamma>0$ and $\gamma<0$ using the joint dataset: a level of confidence of $68.26\%$ ($1-\sigma$) and $95.44\%$  ($2-\sigma$) }\label{fig1}
  \end{center}
\end{figure*}

In this section, we estimate the best-fit values of the cosmological parameters and model parameters using the MCMC algorithm by constraining both FLRW and Bianchi models with recent observational data, including f$\sigma_8(z)$ $+$ BBN $+$ Hubble $+$ BAO $+$ Pantheon$+$ joint dataset. Fig.~\eqref {fig1} includes: a) the 1-$\sigma$ and 2-$\sigma$ posterior distributions of the FLRW brane model parameters in two cases depending on the sign of $\gamma$, and b) the 1-$\sigma$ and 2-$\sigma$ posterior distributions of the Bianchi type-I and V brane model parameters in two cases including $\gamma>0$ and $\gamma<0$. According to b in Fig.~\eqref{fig1}, due to the small value of $\gamma$, the values of all free parameters are approximately the same in both cases $\gamma>0$ and $\gamma<0$. It implies that $\gamma$ is almost negligible in the Friedmann equation of the Bianchi model.

\begin{figure*}
	\begin{center}
  \includegraphics[width=10cm]{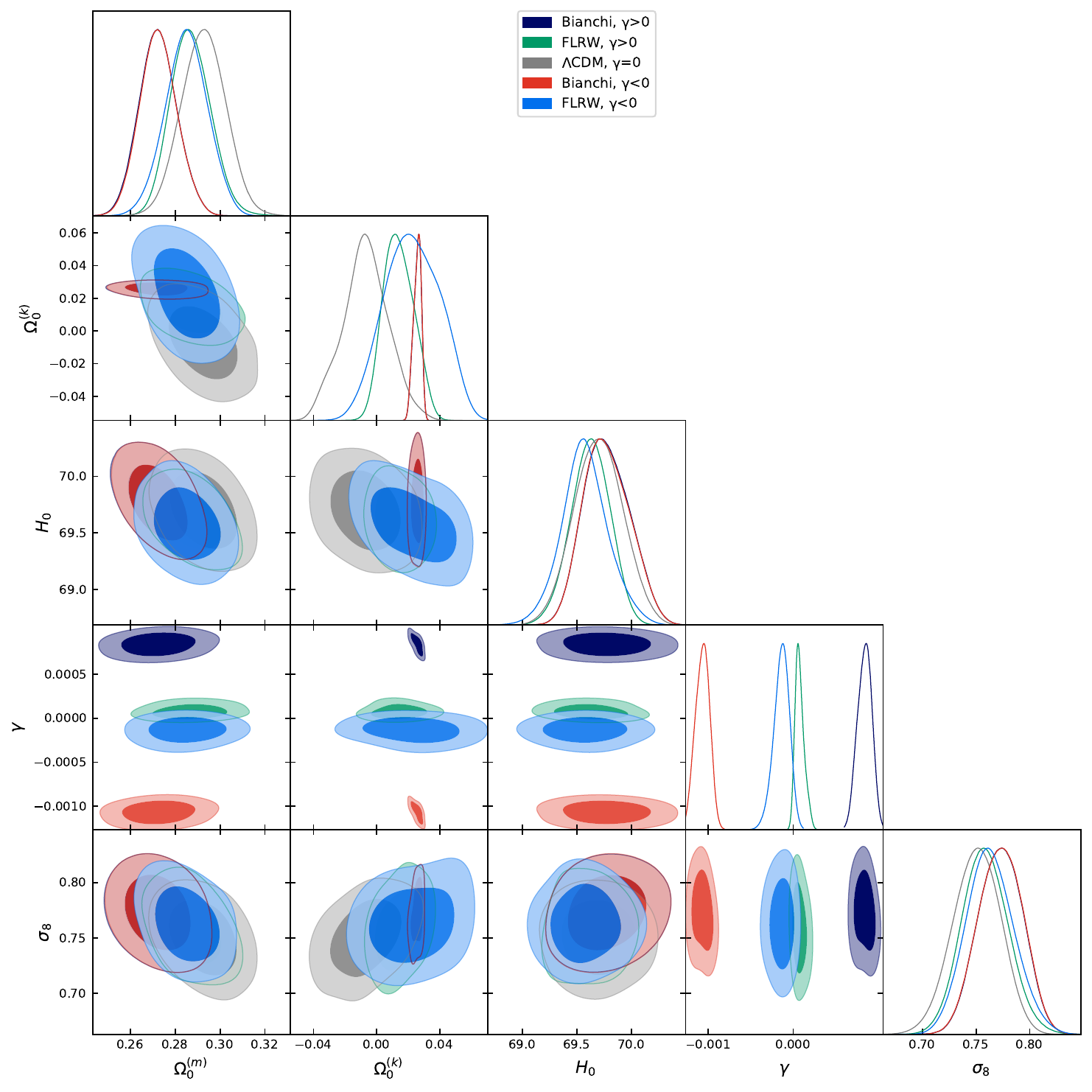}
  \caption{\small  The comparison of the likelihood contours for Bianchi type-I and V model parameters in two cases $\gamma>0$ and $\gamma<0$, FLRW model parameters in two cases $\gamma>0$ and $\gamma<0$, and $\Lambda$CDM$+\Omega^{\text(k)}$ model ($\gamma=0$) using the joint dataset: a level of confidence of $68.26\%$ ($1-\sigma$) and $95.44\%$  ($2-\sigma$)    }\label{fig2}
\end{center}
\end{figure*}

Fig.~\eqref{fig2} presents the comparison of the FLRW and Bianchi type-I and V brane models with the $\Lambda$CDM $+\Omega^{\text(k)}$ model ($\gamma=0$). Utilizing observational data as detailed in the preceding sections, it has been determined that the density parameter related to the metric parameter $\beta$ in the Bianchi model is close to zero. This suggests that the density parameter of this parameter is like the curvature density parameter within Bianchi type-I and V cosmology. Therefore, we display $\Omega^{(\beta)}_0$ with $\Omega^{\text{(k)}}_0$ in Fig.~\eqref{fig2} to compare all models together simultaneously.

Table~\eqref{table1} provides the best-fitting values for the free parameters of the FLRW brane, Bianchi type-I and V brane, and $\Lambda$CDM $+\Omega^{\text{(k)}}$ models. {Acording to Table~\eqref{table1}, in the $\Lambda$CDM $+\Omega^{\text{(k)}}$ model ($\gamma=0$), $\Omega^{\text{(k)}}_0$ have a minimal negative value which presents an open universe inclined to the flat one. In the both cases of FLRW metric ($\gamma>0$ and $\gamma<0$), $\Omega^{\text{(k)}}_0$ is positive which represents a closed universe. In the Bianchi model in both cases $\gamma>0$ and $\gamma<0$, the value of $\Omega^{(\beta)}_0$ is positive which represents the Bianchi type-V model (in the Bianchi type-V model, $\beta=0$).}
 All figures and tables below have been analyzed using the Python package $\texttt{getdist}$~\cite{Lewis:2019xzd}.

\begin{table*}[!htbp]
\begin{center}
\begin{tabular}{|c|c|c|c|c|c|}
 \hline

 Parameter &  $\Lambda$CDM$+\Omega^{\text{(k)}}$, $\gamma=0 $&  FLRW, $\gamma>0$ & FLRW, $\gamma<0$ & Bianchi,  $\gamma>0$ & Bianchi, $\gamma<0$ \\
\hline
{$\Omega^{\text{(m)}}_0 $}  & $0.293^{+0.026}_{-0.025}   $  &  $0.287^{+0.025}_{-0.020}   $& $0.285^{+0.022}_{-0.025}   $& $0.272^{+0.021}_{-0.018}   $& $0.272^{+0.022}_{-0.018}   $ \\

{$\Omega^{\text{(k)}}_0   $} & $-0.007^{+0.039}_{-0.035}  $ & $0.014^{+0.024}_{-0.022}$&
$0.023^{+0.039}_{-0.041}   $ & $0$&
$0$\\

{$\Omega^{(\beta)}_0   $} & $0  $ & $0$&
$0  $ & $0.0258^{+0.0052}_{-0.0063}$&
$0.0259^{+0.0050}_{-0.0062}$\\

{$H_0            $} & $69.70^{+0.56}_{-0.57}     $ & $69.62^{+0.44}_{-0.49}   $&
$69.58^{+0.57}_{-0.57}     $ & $69.77^{+0.52}_{-0.48}     $
&$69.78^{+0.51}_{-0.43}     $\\

{$\gamma         $} & $ 0$ & $0.00008^{+0.00015}_{-0.00011}$ & $-0.0226^{+0.0054}_{-0.0062}$&
$0.00084^{+0.00019}_{-0.00021}$ &
$-0.00107^{+0.00019}_{-0.00020}$\\

{$\sigma_8            $} & $0.750^{+0.053}_{-0.059}   $ & $0.758^{+0.061}_{-0.054}   $ & $0.763^{+0.060}_{-0.053}   $& $0.773^{+0.049}_{-0.051}   $&
$0.772^{+0.050}_{-0.051}   $  \\

{$\Omega^{(\theta)}_0(\times 10^{-5} )$ }&$0$&$0$&$0$&$4.19^{+0.67}_{-0.75}$&$4.17^{+0.91}_{-0.98}$\\
\hline
{\boldmath$\chi^2_{\text{min}}   $} & $821.00 $  & $821.00$  & $820.97$  & $816.05$ & $816.05$ \\
\hline
\end{tabular}
  \caption{The best-fitting values of the free parameters of $\Lambda$CDM+$\Omega^{\text(k)}$ model ($\gamma=0$), FLRW model in two cases $\gamma>0$ and $\gamma<0$ and the Bianchi type-I and V model in two cases $\gamma>0$ and $\gamma<0$  using the $BBN + Pantheon Plus + Hubble + BAO + f\sigma_8(z)$ joint dataset}
    \label{table1}
    \end{center}
\end{table*}
\begin{figure*}[!htbp]
\begin{center}
	\includegraphics[height=4cm,width=8cm]{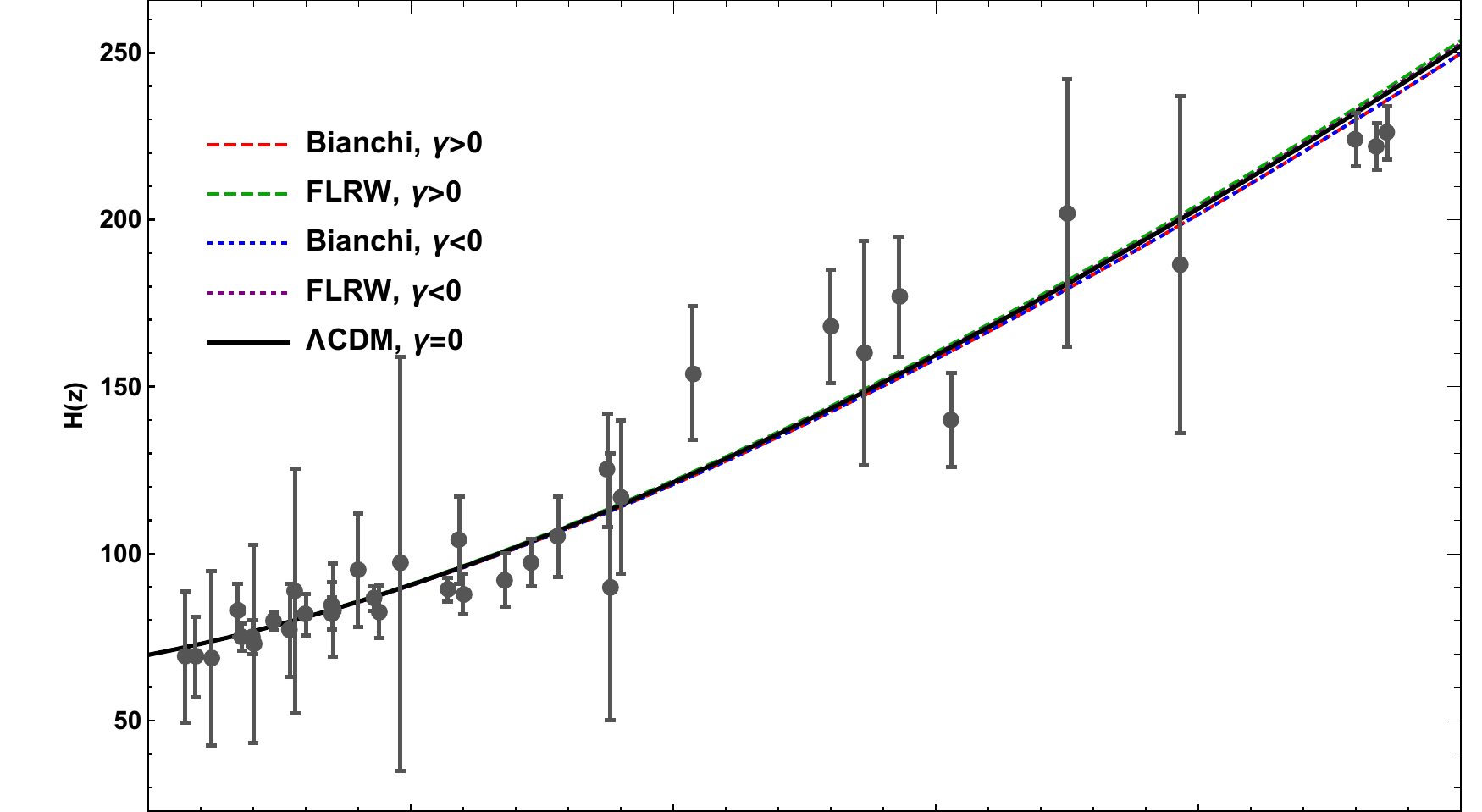}
	\includegraphics[height=4cm,width=8cm]{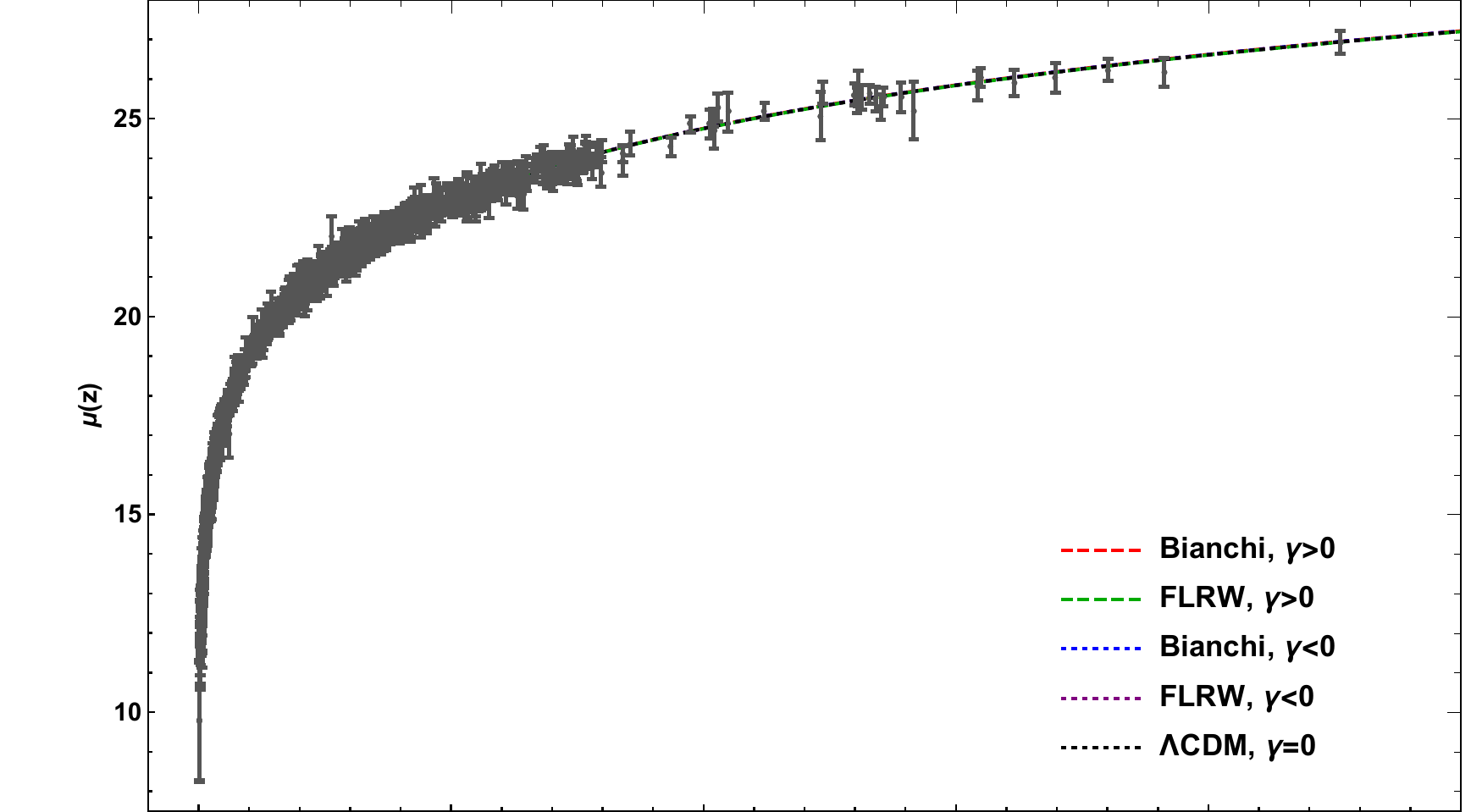}
	\includegraphics[width=8cm]{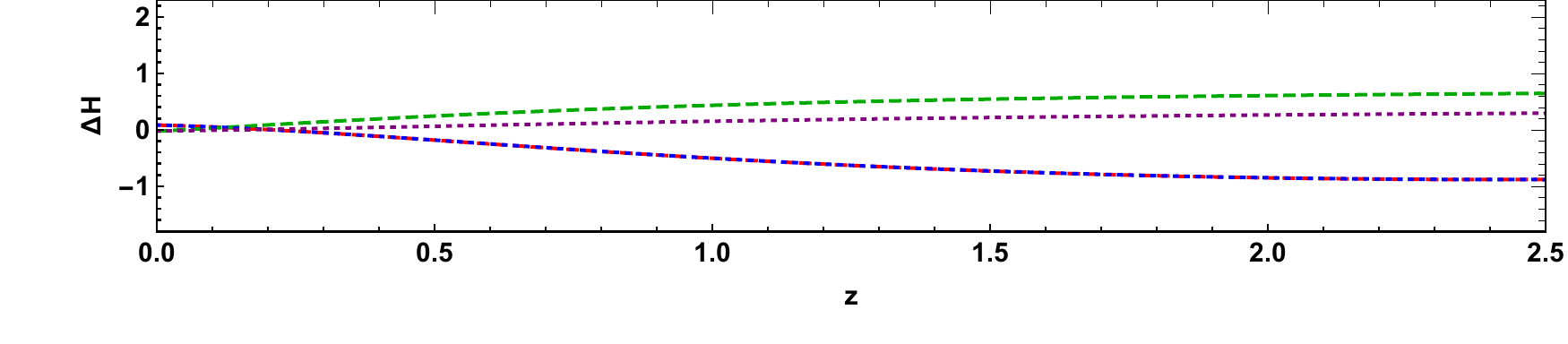}
	\includegraphics[width=8cm]{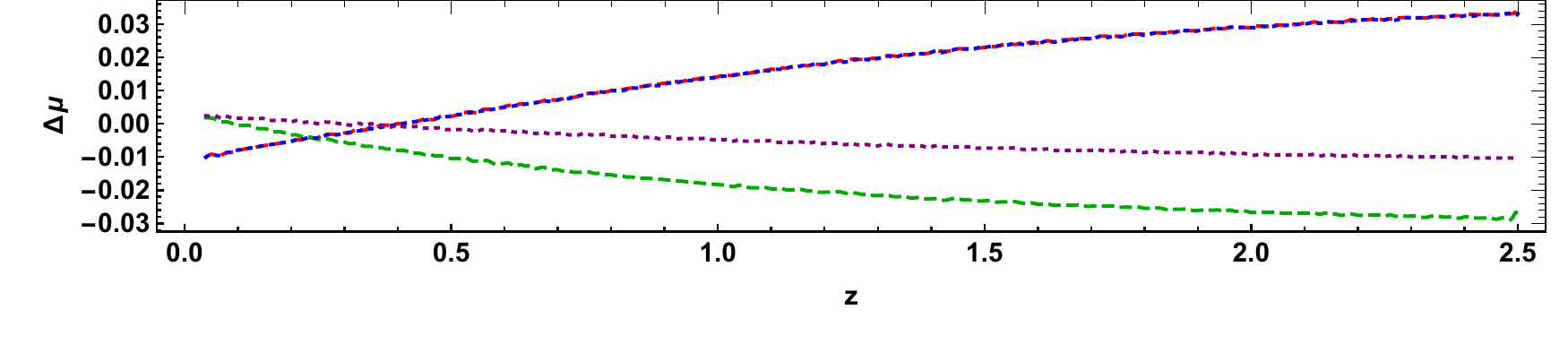}
	\caption{left panels: The Hubble parameter for all mentioned models as a function of redshift was theoretically predicted using the best-fitting values of free parameters presented in Table \ref{table1} in contrast to the observational data derived from cosmic chronometers~\citep{Farooq:2016zwm}. right panels: The evolution of the distance modulus for all mentioned models as a function of redshift in contrast to the observational data derived from \citep{Scolnic:2021amr}. The best-fit values from Tables \ref{table1} are utilized in all the models.}
	\label{fig3}
\end{center}
\end{figure*}

\begin{figure*}[!htbp]
\begin{center}
	\includegraphics[height=4cm,width=8cm]{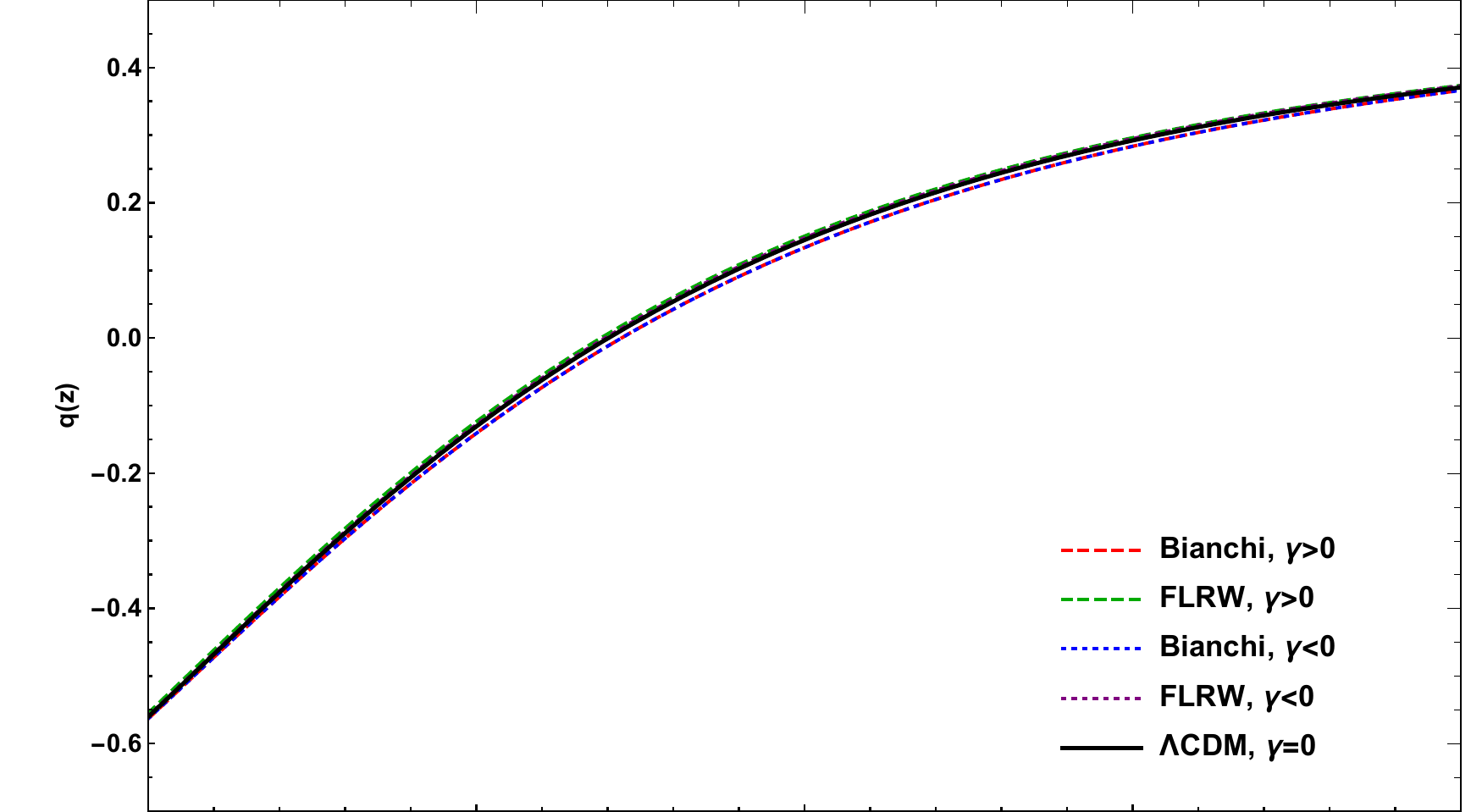}
	\includegraphics[height=4cm,width=8cm]{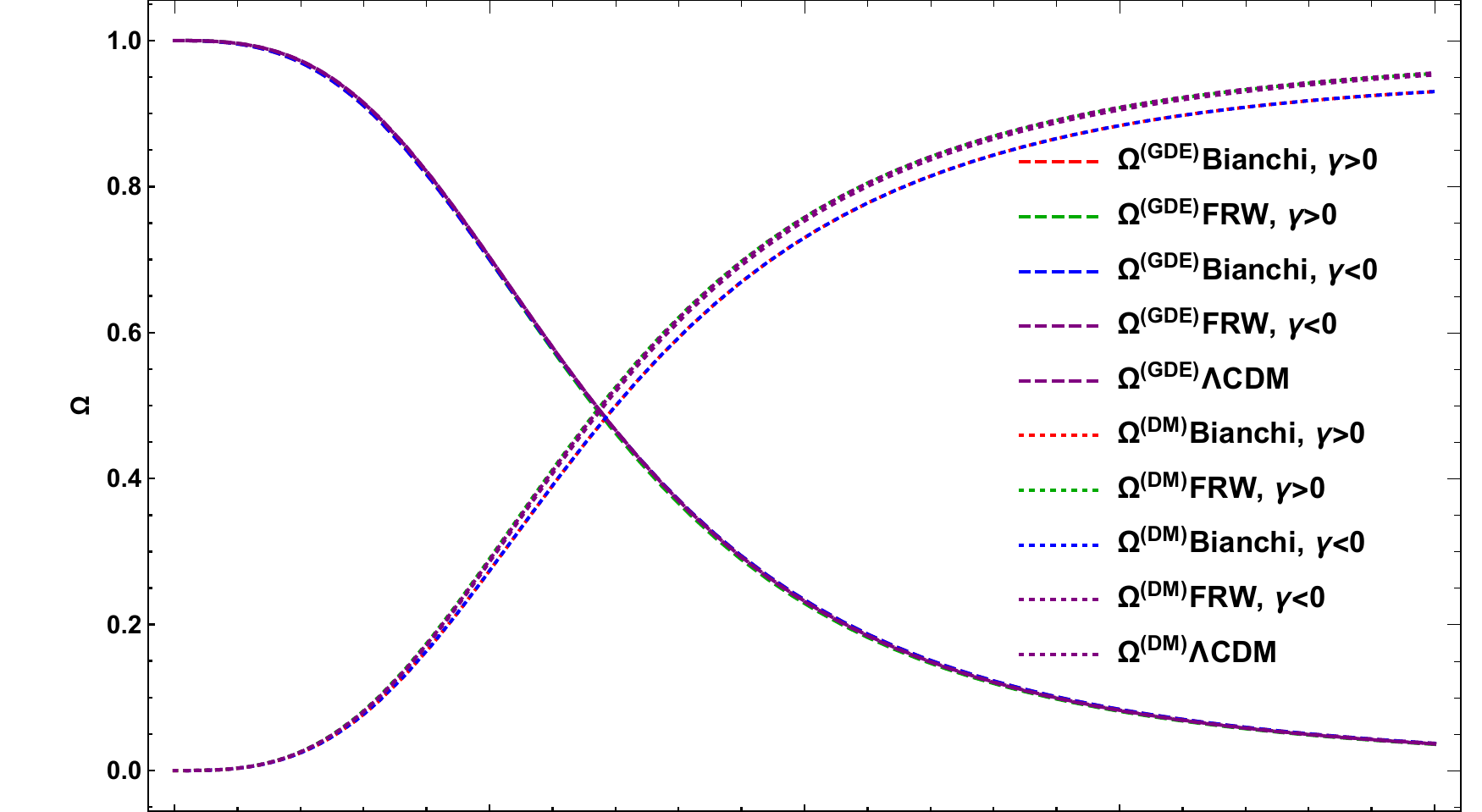}
	\includegraphics[width=8cm]{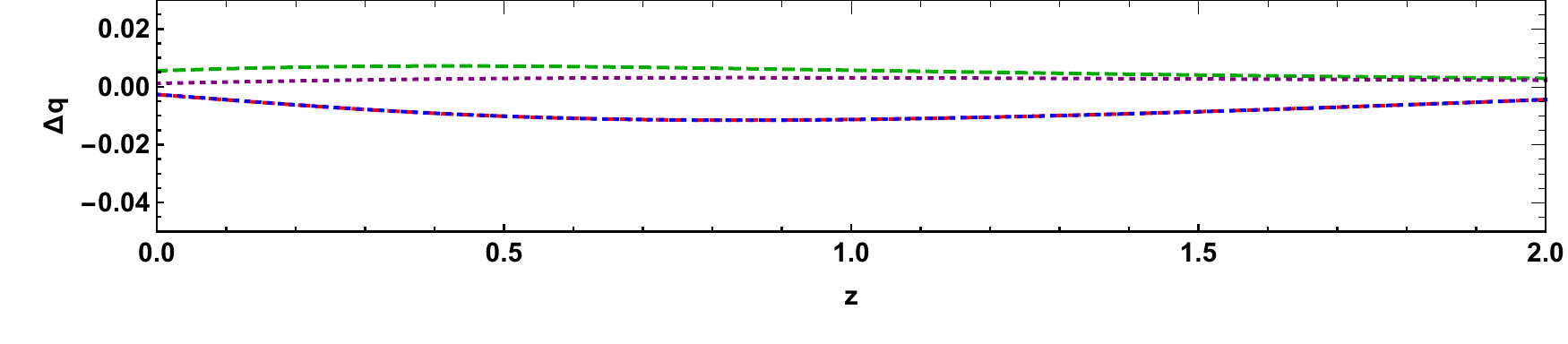}
	\includegraphics[width=8cm]{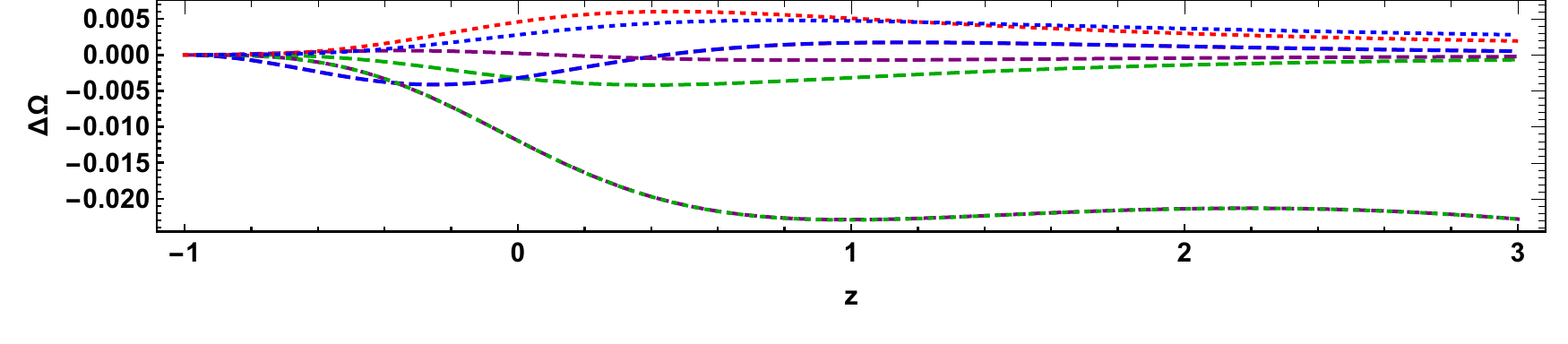}
	\caption{left panels: The evolution of the deceleration parameter for all mentioned models as a function of redshift. The best-fit values from Tables \ref{table1} are utilized in all the models. right panels: The evolution of the density parameters for all mentioned models as a function of redshift. The best-fit values from Tables \ref{table1} are utilized in all the models.}
	\label{fig4}
\end{center}
\end{figure*}
\begin{figure}[!htbp]
\begin{center}
	\includegraphics[height=4cm,width=8cm]{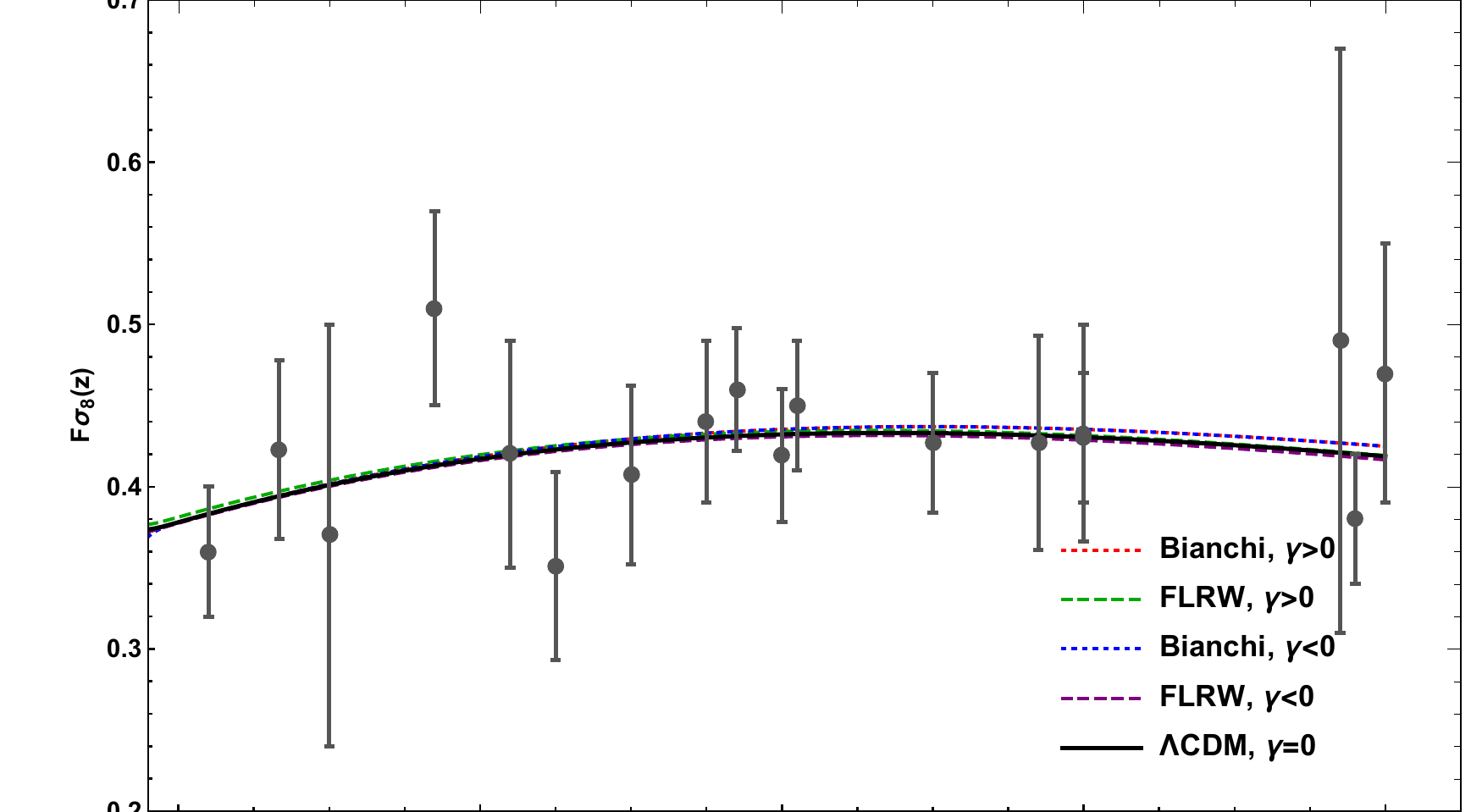}
	
	\includegraphics[width=8cm]{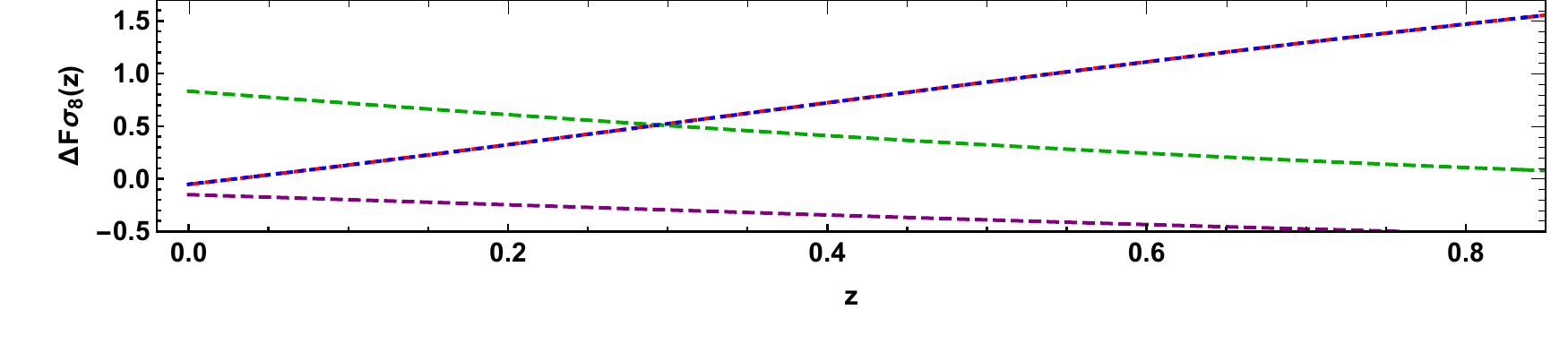}
	
	\caption{The f$\sigma_8(z)$ quantity for all mentioned models as a function of redshift using the best-fitting values of free parameters presented in Table \ref{table1} in contrast to the observational data derived from \citep{Nesseris:2017vor}}
	\label{fig5}
\end{center}
\end{figure}

\begin{table*}[!htbp]
\begin{center}
\begin{tabular}{|c|c|c|c|c|c|}
 \hline
 Parameter &  $\Lambda$CDM$+\Omega^{\text{(k)}}$, $\gamma=0 $&  FLRW, $\gamma>0$ & FLRW, $\gamma<0$ & Bianchi,  $\gamma>0$ & Bianchi, $\gamma<0$  \\
\hline
{$q_0   $} & $-0.56$  & $ -0.55$& $-0.55 $& $-0.56$ &$ -0.56$\\
{$t_0           $} & $13.66$ &$13.61$ & $13.64$&$13.53$ & $13.53$\\
{$\omega           $} &$-1$  &$-1$ &$ -0.99$&$ -1.00002$& $-0.99$\\
{$\Omega^{\text{(GDE)}}_0             $} &  $0.703414$  & $0.700194$ & $0.703624$&$0.700201$ &$ 0.700201$\\
{$z_t   $} & $0.70$ &$0.68$ &$0.69$ &$0.72$ & $0.72$\\
\hline
\end{tabular}
    \caption{the values of age of the universe, deceleration parameter at the present epoch, the equation of state parameter, the corresponding transition redshift of the Bianchi type-I and V model parameters in two cases $\gamma>0$ and $\gamma<0$, FLRW model in two cases $\gamma>0$ and $\gamma<0$, and $\Lambda$CDM model ($\gamma=0$) using the $BBN + Pantheon Plus + Hubble + BAO + f\sigma_8(z)$ joint dataset}
    \label{table2}
 \end{center}
\end{table*}
\begin{table*}[!htbp]
\begin{center}
\begin{tabular}{|c|c|c|c|c|c|}
 \hline
 Parameter &  $\Lambda$CDM$+\Omega^{\text{(k)}}$, $\gamma=0 $&  FLRW, $\gamma>0$ & FLRW, $\gamma<0$ & Bianchi,  $\gamma>0$ & Bianchi, $\gamma<0$  \\
\hline
{AIC  } & $831.00$  & $ 833.00$& $832.97$& $830.04$ &$ 830.04$\\
{BIC  } & $858.36$ &$865.83$ & $865.80$ & $868.36$& $868.36$\\
{DIC  } &$830.77$  &$828.56$ &$ 830.13$&$ 824.34$& $824.592$\\
\hline
\end{tabular}
    \caption{Statistical criteria of models are derived through background and perturbation observable data.}
    \label{table3}
 \end{center}
\end{table*}

Our second step involves predicting the evolution of cosmological parameters such as the Hubble parameter $H(z)$, the deceleration parameter $q(z)$, the distance modulus $\mu(z)$, the equation of state parameter $\omega(z)$, the density parameters of both
matter $\Omega^{\text{(m)}}$ and geometric dark energy $\Omega^{\text{(GDE)}}$, and the growth rate of structures $f\sigma_8(z)$ as functions of redshift, based on the best fitting values of the free parameters of both models indicated in Table~\eqref{table1}.\\ Figs.~\eqref{fig3}, \eqref{fig4}, and \eqref{fig5} depict the evolution of these functions. In Fig.~\eqref{fig3}, we plot the Hubble parameter as a function of redshift for all models. We show the relative difference between the Hubble parameters in considered models with $\Lambda$CDM model, $\Delta H(z)=(\frac{H_{\text{model}}-H_{\Lambda CDM}}{H_{\Lambda CDM}})\cross 100$ in the bottom right panel of Fig.~\eqref{fig3}. All models behave like $\Lambda$CDM in small redshifts ($\Delta H=0$). However, at larger redshifts, for the FLRW model with $\gamma>0$, $\Delta H$ is positive, while for the FLRW model with $\gamma<0$, $\Delta H=0$. For the Bianchi model, $\Delta H$ is negative in both cases $\gamma>0$ and $\gamma<0$.

In the right panels of Fig.~\eqref{fig3},  we present the evolution of the distance modulus and the relative difference of distance modulus $\Delta \mu(z)=(\frac{\mu_{\text{model}}-\mu_{\Lambda CDM}}{\mu_{\Lambda CDM}})\cross 100$ for all models. As observed, the behavior of the FLRW and Bianchi models is opposite. At small redshifts, $\Delta \mu(z)$) is positive for the FLRW models but negative for the Bianchi models.

In the right panels of Fig.~\eqref{fig4}, we plot the evolution of the deceleration parameter and $\Delta q=q_{\text{model}}-q_{\Lambda CDM}$. For both cases of the Bianchi model, $\Delta q <0$, and the corresponding transition redshift in both cases is $z_t=0.72$, which is higher than the corresponding transition redshift in $\Lambda$CDM model $z_t=0.70$. In the FLRW models, the deceleration parameter behaves like the $\Lambda$CDM model and $\Delta q \simeq 0$, and the corresponding transition redshift is $z_t=0.69$. In the left panels of Fig.~\eqref{fig4}, we show the evolution of the density parameters of both matter $\Omega^{\text{(m)}}$ and geometric dark energy $\Omega^{\text{(GDE)}}$  and $\Delta \Omega={\Omega}_{\text{model}}-{\Omega}_{\Lambda CDM}$ for all models. In all cases, models at the large redshifts tend to be in the matter-dominated era but in small redshifts, they tend to be in the geometric dark energy-dominated era.

In Fig.~\eqref{fig5}, we plot the evolution of the growth rate of matter fluctuations for all models including $\Lambda$CDM, FLRW, and Bianchi. At the bottom of Fig.~\eqref{fig5}, we have demonstrated $\Delta f\sigma_8(z)=(\frac{{(f\sigma_8)}_{\text{model}}-{(f\sigma_8)}_{\Lambda CDM}}{{(f\sigma_8)}_{\Lambda CDM}})\cross 100$. It is evident that $\Delta f\sigma_8(z)$ increases with redshift in both Bianchi models. $\Delta f\sigma_8(z)$ in the FLRW model with $\gamma>0$ in small redshifts is positive and tends to be zero at large redshifts. In the FLRW model with a negative $\gamma$, the quantity $\Delta f\sigma_8(z)$ decreases from zero to negative values with increasing redshift.

We report the values of the universe's age, the deceleration parameter at the present epoch, the equation of state parameter, and the corresponding transition redshift in all mentioned models in Table~\eqref{table2}. In all cases, the universe's age has become smaller than the standard scenario, according to Figs. \eqref{fig3} to \eqref{fig5}, the models exhibit no significant differences in behavior among these functions.\\
Generally, a method for determining if a model is numerically close to observational data is necessary to select the most appropriate model. In this study, we make use of several statistical criteria, including the Akaike information criterion (AIC), the Bayes factor or Bayesian Information Criterion (BIC), and the Deviance Information Criterion (DIC)~\cite{Capozziello:2011tj}. These statistical criteria are defined in~\cite{Kass:1995loi, Spiegelhalter:2002yvw, whitehead2007selection, liddle2007information,  Rezaei:2021qpq}. 
\begin{equation}\label{eq50}
\begin{split}
AIC&=\chi^2_{\text{min}}+2M,\\
BIC&=\chi^2_{\text{min}}+M\ln{N},\\
DIC&=2\overline{\chi^2_{\text{tot}}(p)}-\chi^2_{\text{tot}}(\overline{p}),
\end{split}
\end{equation}
which N represents the total number of data points, and M represents the number of free parameters.
We summarise the statistical results of the mentioned models by using BBN, Pantheon$+$, Hubble, BAO, and f$\sigma_8(z)$ joint dataset in Table (\ref{table3}).
By choosing $\Lambda$CDM$+\Omega^{\text{(k)}}$ as a reference model, we calculate the difference $\Delta$AIC, $\Delta$BIC and $\Delta$DIC of Bianchi type-I and V and FLRW brane models with $\Lambda$CDM$+\Omega^{\text{(k)}}$  model.

According to the $\Delta$AIC criteria, the value of $\Delta \text{AIC} \leq 2$ indicates that the model is sufficiently supported compared to the reference model, In contrast, $4 \leq \Delta \text{AIC} \leq 7$ suggests less support, and $\Delta \text{AIC} \geq 10$ signifies lack of support~\cite{whitehead2007selection}. The results in Table (\ref{table3}) indicate good compatibility of all considered models with observational data when compared to $\Lambda$CDM$+\Omega^{\text{(k)}}$ ($\Delta \text{AIC} < 2$).

Similarly, for the Bayesian Information Criterion (BIC), a difference falling within $0 < \Delta \text{BIC} \leq 2$ is considered weak evidence, $2 < \Delta \text{BIC} \leq 6$ is positive evidence, $6 < \Delta \text{BIC} \leq 10$ is strong evidence and any difference exceeding 10 is considered strong evidence against the model with the higher BIC~\cite{Kass:1995loi}. Therefore, there is strong positive evidence against all cases.

Finally, if $\Delta \text{DIC} \leq 2$, the model is statistically compatible with the reference model, When $2 < \Delta \text{DIC} < 6$, it signifies moderate tension between the two models, whereas $\Delta \text{DIC} \geq 10$ indicates significant tension~\cite{anagnostopoulos2020observational}. From $\Delta$DIC, we can conclude that the data supports the FLRW brane model with $\gamma<0$ more than others. Finally, we can estimate the upper limit for the relative time variation of the gravitational constant $(\frac{\dot G_N}{G_N})$.
We found $\gamma =-0.0226^{+0.0054}_{-0.0062}$ and $H_0=69.58^{+0.57}_{-0.57} $ in the FLRW model with $\gamma<0$. Substituting these values into equation~\eqref{eq10}, we found $ \frac{\dot G_N}{G_N} =(-1.4^{+0.36}_{-0.42})\times 10^{-12}$. This value of $\frac{\dot G_N}{G_N}$ is compatible with other findings in Refs.~\citep{thorsett1996gravitational, jofre2006constraining, krastev2007constraining, ray2007dark, anderson2015measurements, zhao2018constraining}.

\section{Conclusion}\label{concl}
This study assesses the compatibility of the CEG model, extensively discussed in Refs.~\citep{Jalalzadeh:2013wza, Jalalzadeh:2023upb}, with the latest observable data such as BBN, Pantheon$+$, Hubble, BAO, and f$\sigma_8(z)$. Our study is conducted in three steps.
Initially, the FLRW and Bianchi type-I and V brane models are constrained with the mentioned joint dataset at the background and perturbation levels. The MCMC method is employed to ascertain the best-fitting values of the free parameters in both models, elucidated in Table~\ref{table1}.

Subsequently, we scrutinize the behavior of various cosmological parameters such as the Hubble parameter $H(z)$, the deceleration parameter $q(z)$, the distance modulus $\mu(z)$, the equation of state parameter $\omega(z)$, and the density parameters of matter $\Omega^{\text{(M)}}$ and geometric dark energy $\Omega^{\text{(GDE)}}$. Additionally, we examine the growth rate of structures f$\sigma_8(z)$ as functions of redshift. This analysis is predicated on the best-fitting values of the independent parameters of the FLRW and Bianchi type-I and V brane models within the framework of CEG, depicted in Figs.~\eqref{fig3}, \eqref{fig4}, and \eqref{fig5}.\\
Lastly, we utilize statistical criteria such as AIC, BIC, and DIC to gauge the compatibility of CEG with the $\Lambda$CDM$+\Omega^{\text{(k)}}$ model. The outcomes of these statistical analyses are presented in Table~\ref{table3}. Notably, the study on $\Delta DIC$ suggests that the FLRW model with $\gamma<0$ and the $\Lambda$CDM$+\Omega^{\text{(k)}}$ model exhibit equally good fits to the observational data compared to the other mentioned models.

{By conducting a thorough comparison between our computational models and empirical observational data, we are able to ascertain the optimal values for various cosmological parameters. Within the framework of the FLRW model, these parameter values are contingent upon the nature of the sign of $\gamma$, which essentially denotes the rate of change of the gravitational constant in units of the Hubble time. Specifically, when $\gamma >0$, the derived value is $\gamma=0.00008^{+0.00015}_{-0.00011}$, alongside $\Omega^{\text{(k)}}_0 = 0.014^{+0.024}_{-0.022}$, whereas for $\gamma < 0$, the corresponding figures are $\gamma = -0.0226^{+0.0054}_{-0.0062}$ and $\Omega^{\text{(k)}}_0 = 0.023^{+0.039}_{-0.041}$. It is essential to highlight that in both scenarios, it holds true that $\Omega^{\text{(k)}}_0$ is greater than zero, thereby indicating the presence of a closed universe. Similarly, in the context of the Bianchi type-V brane model, the parameter values exhibit variability based on the sign of $\gamma$. This variation results in $\gamma = 0.00084^{+0.00019}_{-0.00021}$, $\Omega^{(\beta)}_0=0.0258^{+0.0052}_{-0.0063}$, and $\Omega^{\theta}_0(\times 10^{-5}) = 4.19^{+0.67}_{-0.75}$ (corresponding to the density parameter of stiff matter) for cases where $\gamma > 0$. Conversely, for instances where $\gamma < 0$, the parameter values are $\gamma = -0.00107^{+0.00019}_{-0.00020}$, $\Omega^{(\beta)}_0 = 0.0259^{+0.0050}_{-0.0062}$, and $\Omega^{\theta}_0(\times 10^{-5}) = 4.17^{+0.91}_{-0.98}$. Importantly, in both scenarios, the condition $\Omega^{(\beta)}_0 > 0$ holds, thereby characterizing the nature of the Bianchi type-V model.}

\section*{Acknowledgments}
S.J. acknowledges financial support from the National Council for Scientific and Technological Development, Brazil--CNPq, Grant no. 308131/2022-3.
\section*{Data availability}
All data used in this work are already publicly available.
\bibliography{Dark}
\end{document}